# Subcarrier Multiplexing for Parallel Data Transmission in Indoor Visible Light Communication Systems

Safwan Hafeedh Younus[1], Aubida A. Al-Hameed[1], Ahmed Taha Hussein[1], Mohammed T. Alresheedi[2] and Jaafar M. H. Elmirghani[1]
[1]School of Electronic and Electrical Engineering, University of Leeds, LS2 9JT, United Kingdom
[2]Department of Electrical Engineering, King Saud University, Riyadh, Kingdom of Saudi Arabia
elshy@leeds.ac.uk, elaawj@leeds.ac.uk, asdftaha@yahoo.com, malresheedi@ksu.edu.sa, j.m.h.elmirghani@leeds.ac.uk

*Abstract:*— This paper presents an indoor visible light communication (VLC) system in conjunction with an imaging receiver with parallel data transmission (spatial multiplexing) to decrease the effects of inter-symbol interference (ISI). To distinguish between light units (transmitters) and to match the light units used to convey the data with the pixels of the imaging receiver, we propose the use of subcarrier multiplexing (SCM) tones. Each light unit transmission is multiplexed with a unique tone. At the receiver, a SCM tone decision system is utilized to measure the power level of each SCM tone and consequently associate each pixel with a light unit. In addition, the level of co-channel interference (CCI) between light units is estimated using the SCM tones. Our proposed system is examined in two indoor environments taking into account reflective components (first and second order reflections). The results show that this system has the potential to achieve an aggregate data rate of 8 Gb/s with a bit error rate (BER) of $10^{-6}$ for each light unit, using simple on-off-keying (OOK).

## I. INTRODUCTION

Currently, radio frequencies (RFs) are utilized to transport the data in wireless systems. However, it is generally accepted that there is a shortage in the RF spectrum, so RF may not be able to continue to meet the increasing traffic demands [1]. Recent research by Cisco has shown that between 2016 and 2021 mobile data traffic will grow sevenfold [2]. Thus, many researchers have suggested visible light communication (VLC) systems as an alternative or a complementary technique to overcome the deficit in the RF spectrum. VLC systems offer many significant benefits over RF systems, such as virtually unlimited bandwidth and unlicensed spectrum, high security, and the simplicity and the low cost of the transmitter and detector devices [3]. However, VLC systems are based on line-of-sight (LOS) connections; hence, the absence of the LOS components between the receiver and the transmitter greatly impact the performance of the VLC system. In addition, due to light not passing through opaque objects, access points need to be interconnected via a wire / fibre that needs to be installed to connect rooms. Single color or multi-color light emitting diodes (LEDs) have been considered as sources of illumination and data communication in indoor VLC systems. However, LEDs have a slow response (normally tenth of megahertz for the single colour LED and hundreds megahertz for the multi-colour LED) [1], [4], laser diodes (LDs) have been suggested to realize multi-gigabit data transmission rates for indoor VLC systems [5]-[8]. Comparing LDs with LEDs, LDs offer higher modulation speeds and better efficiency for the electrical to optical conversion, which makes them better candidates as transmitters at high data rates [9].

VLC systems are still under improvement and many obstacles encounter them, particularly in terms of achieving high data rates while utilizing the VLC spectrum efficiently. Many directions have been suggested to improve the data rate of the indoor VLC systems. One of these directions is the use of beam steering in optical wireless communication systems to realize multi Gb/s wireless communication indoor [10]-[12]. However, this can lead to increased design complexity, as an extra optical component (i.e., spatial light modulator or computer generated holograms) should be added at the transmitter. Multiple input multiple output (MIMO) techniques are another direction that can be followed to enhance the performance of the indoor optical wireless communication systems [4]. When used in indoor VLC systems, these MIMO systems have to ensure that an acceptable illumination level is achieved. Thus, arrays of LEDs or LDs are used to ensure a sufficient level of illumination in the indoor environment with potentially higher transmission data rates. This makes MIMO systems very attractive for indoor VLC systems to enhance the achievable data rate [13]. MIMO techniques have been investigated widely in indoor VLC systems. An indoor VLC MIMO system with imaging receiver and non-imaging receiver was considered in [4]. It was shown that based on the location of the receiver, a lens located at a good place between the transmitters and the receivers can de-correlate the matrix of the MIMO channel. The work in [14] described spatial modulation in a MIMO indoor VLC system. It showed that interference between transmitters can be avoided by activating just one transmitter at any time. An indoor VLC system based on $4 \times 9$ MIMO was shown to achieve a data rate of 1 Gb/s over 1 m, where the equalizer was implemented in both the receiver and the transmitter to extend the bandwidth of the system [15]. A 50 Mb/s data rate was achieved experimentally over a distance of 2 m for an indoor $4 \times 4$ MIMO VLC system [16]. Advanced receivers were proposed to reduce the correlation between the VLC-MIMO channels in [17]-[19]. An angle diversity receiver (ADR) was suggested in [17] to change the direction / angle of each photodetector to a different direction, while in [18] a prism array receiver (PAR) was employed on top of each photodetector to direct the light into a specific orientation to reduce the correlation between the VLC-MIMO channels. In [19], angle-added mirror diversity receiver (AMDR) was proposed to reduce the effect of the correlation between the VLC-MIMO channels.

This paper introduces a SCM parallel data transmission indoor VLC system. We use LDs, which have modulation bandwidth higher than LEDs, as the source of lighting and communication. This system uses parallel data transmission and utilizes an imaging receiver. Each pixel of the imaging receiver is considered as a single photodetector that has a narrow field-of-view (FOV). To distinguish between the light units, the transmission from each light unit was multiplexed with a unique subcarrier multiplexing (SCM) tone. These unmodulated tones are used at the beginning of the communication to set up the connection between the light units and the optical receiver. The SCM tones were used to calculate the level of crosstalk between wavelength division multiplexing (WDM) channels [20] and were utilized in indoor VLC positioning systems [21]. Here, the SCM tones are used to give an ID to each light unit (transmitter), to help find the light units that have a good channel for high data rate transmission and match these light units with pixel(s) of the imaging receiver. The SCM tones were also used to calculate the co-channel interference (CCI) level at each pixel of the imaging receiver. To the best of our knowledge, this is the first time that SCM tones have been suggested to support parallel data transmission in an indoor VLC system. In order to attain a reliable connection between the optical receiver and the transmitters, each light unit transmits a different data stream with a target bit error rate (BER) = $10^{-6}$. Based on inter-symbol interference (ISI) and the level of CCI, each light unit sends a different data stream at a different data rate. Our proposed system was examined in two different room scenarios: an unfurnished room and a furnished room tacking into account the effects of the diffuse reflections (first order and second order reflections). The realistic (furnished) room represents a small office with a door, mini cubicles, windows, bookshelves and furniture. The difficulty associated with both environments is to create LOS components between the receiver and the transmitter at all the optical receiver's potential locations on the communication floor of the suggested rooms. The proposed VLC system offers an aggregate data rate of 8 Gb/s over the entire communication floor of the proposed rooms with simple on-off-keying (OOK) modulation, as the results show.

The paper is organized as follows: The setup of the rooms and configurations of the RYGB LDs-light units are described in Section II. Section III introduces the receiver configurations. Section IV presents the proposed system's design. The results of the simulation of the system in the proposed rooms are given in Section V and conclusions are drawn in Section VI.

## II. VLC ROOMS SETUP

In this paper, we considered two room configurations for the analysis: an unfinished room (room A) and a furnished room (room B). The unfinished room does not have doors and windows while the realistic room is a small office with windows, a door, physical partitions and bookshelves as can be seen Figs. 1 (a) and (b), respectively. The height of room A and room B is 3 m while the area of the rooms is 4 m × 8 m (width × length). The floor of rooms A and B has a reflection coefficient of 0.3 whereas the ceiling and the walls have a reflection coefficients of 0.8 [22]. Room B is a small office that has three large windows, a door, bookshelves, chairs, furniture and small cubicle offices as depicted in Fig. 1 (b). The door and windows have reflection coefficients of zero (no signals are reflected from them). Additionally, two walls in room B (wall $x = 4$ m and wall $y = 8$ m) are covered with bookshelves and filling cabinets as shown in Fig. 1 (b). These two walls have a reflection coefficient of 0.4 [23]. The barriers of the small offices are supposed to either absorb or block the optical signals. In addition, desks, chairs and tables inside room B have reflection coefficients of 0.3, which is similar to the roof. In room B, shadowing is created by the physical barriers and objects that have low reflection coefficients, and this increases the complexity in room B. The reflecting surfaces (floor, walls and ceiling) of the rooms are assumed to be Lambertian reflectors as plaster walls are roughly Lambertian reflectors [22]. A ray tracing algorithm is utilized to evaluate the reflections. Therefore, the rooms are split into a number of equal square-shaped surface elements. These surface elements have a reflection coefficient of $\rho$ and an area of $d_A$. Each one of these surface elements is assumed to behave as Lambertian emitter with $n = 1$ ($n$ denotes the order of the Lambertain emission). High resolution results can be obtained when the area of the small elements is very small. However, when the element gets smaller this increases the computation time exponentially. Therefore, to keep the time of the computations within a reasonable time, in the first reflections, the size of the surface elements is chosen to be 5 cm × 5 cm while in the second reflections, it is selected to be 20 cm × 20 cm.

In this paper, we take into account reflections up to the second order for the simulation. This is due to the fact that second order reflections have a large effect on the performance of VLC systems when operating at a high data rate. Most of the optical power that reaches to the receiver is within first order and second order reflections. In higher order reflections (third and higher), the received optical signals are significantly attenuated [24]. Hence, we considered only first order reflections and second order reflections.

LDs are utilized rather than LEDs to enable high data rates while using simple OOK modulation. Compared with LEDs, LDs offer higher output powers, which leads to a high lumen output, and LDs are much brighter than LEDs [25]. Experimental tests have shown that multicolour LDs can be used for lighting without any risk on the human eye [26]. Hence, to provide an acceptable white colour in the room and to make it safe for the human eye, we consider the four colours (RYGB) LDs that were utilized in [26]. It should be noted that the RYGB LDs of each light unit are modulated by the same data (the four LDs carry the same data). However, each RYGB LDs-light unit is used to send a different data stream. To get a suitable level of lighting in the room that meets the ISO and European lighting requirements [27], eight RYGB LDs-light units are utilized for lighting. These RYGB LDs-light units are installed on the ceiling of the rooms (3 m over the room's floor) as can be seen in Fig. 1 (a). Each RYGB LDs-light unit has nine RYGB LDs (3 × 3) with a spacing of 3 cm [8]. The azimuth and the elevation of the RYGB LDs-light units are set to 0º and 90º, respectively. Due to the use of a diffuser, the output of the RYGB LDs is assumed to have a Lambertian radiation pattern. Calculations of the level of the lighting can be found in [8] and [28].

## III. IMAGING RECEIVER CONFIGURATION

An imaging receiver was utilized in this work rather than a wide FOV receiver. Several benefits can be achieved by using the imaging receiver such as i) one concentrator is used for all photodetectors and this leads to decrease in the receiver's size and cost and ii) many pixels are used with one planner array for photo-detectors [29]. Moreover, imaging receivers reduce the delay spread because of a small range of rays received by each pixel. This improves the bandwidth of the VLC channel. In this imaging receiver, pixels are laid out to form one detector divided into a number of small elements. The area of these elements are equal and they have rectangular shapes with no spaces among them [30]. The area of each pixel can be obtained by dividing the area of the photo-detector by the number of the pixels. In this paper, the imaging receiver has 288 pixels. A lens is utilized as a concentrator to gather and focus the optical signal from a large area down to a smaller one (see Fig. 2). In the simulation, we used the same lens that was utilized in [29]. This lens has an entrance diameter of 3

The imaging concentrator's transmission factor is given as [29]:

$$T_c(\delta) = -0.1982\delta^2 + 0.0425\delta + 0.8778 \qquad (2)$$

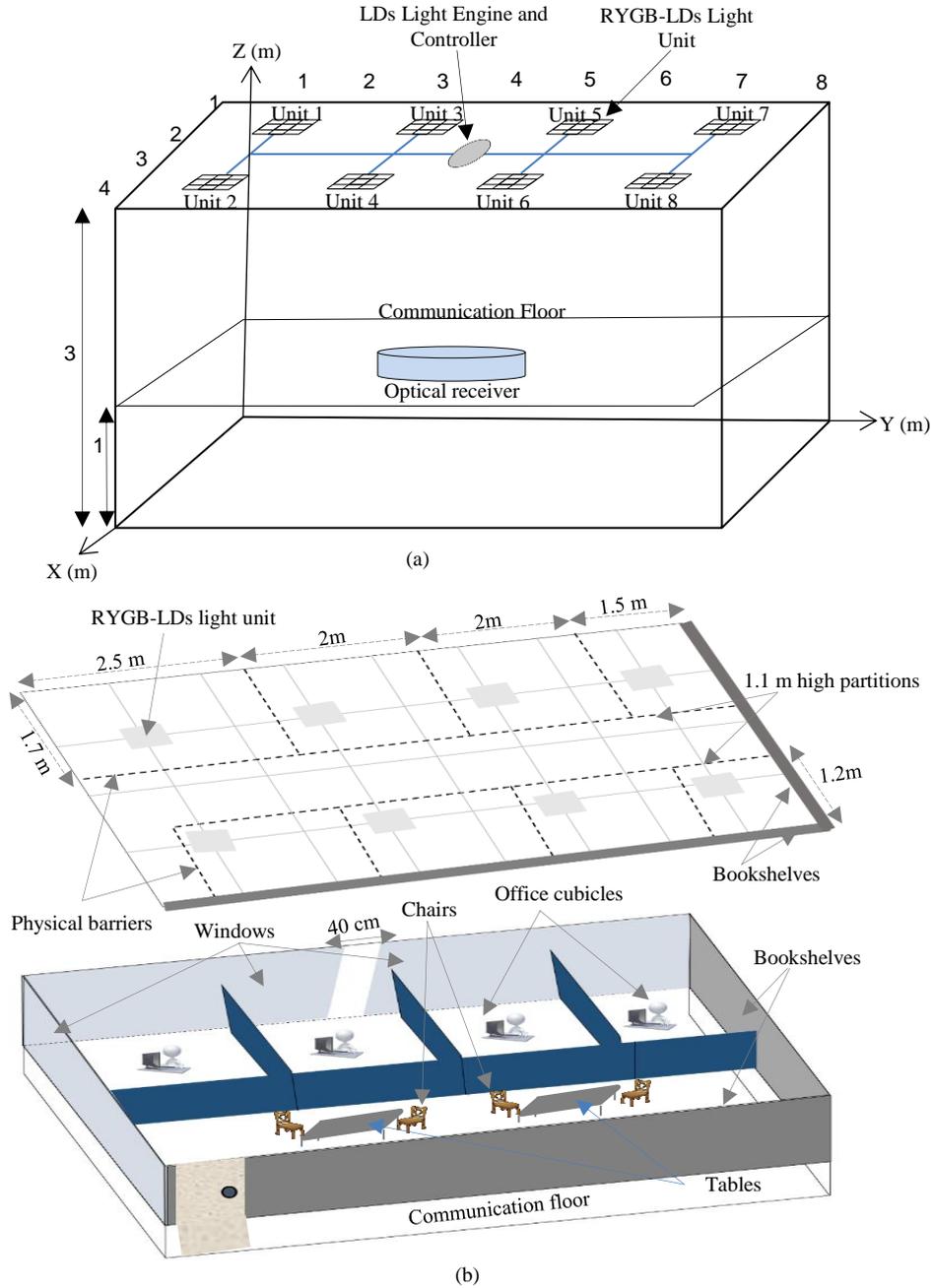

Fig 1: Rooms structure: (a) the unfurnished room and (b) the furnished room.

cm. Consequently, the area of the entrance of this lens is $A = \frac{9\pi}{4}$ cm² with an exit area $A' = \frac{A\sin^2(\psi)}{N^2}$, here $N$ is the lens refractive index ($N = 1.7$) and $\psi$ is the lens receiving angle ($\psi < 90°$). The gain of the lens is:

$$g(\psi) = \frac{N^2}{\sin^2(\psi)} \qquad (1)$$

here $\delta$ is the angle of incidence, which is measured in radians. The FOV of the concentrator is selected to be 65°, which enables the imaging receiver to see the whole ceiling of the room when the optical receiver is located at the room centre. Hence, the FOV of each photodetector is 4.7°. In addition, the detector array's size is chosen to be equal to the concentrator's exit area. Hence, the area of the photodetector is 2 cm² and the pixel's area is 0.694 mm². The azimuth and the elevation of the imaging receiver (for all pixels) are set to 0° and 90°, respectively. It should be noted that each pixel of the imaging receiver electrically amplifies the photo-currents received independently as illustrated in Fig. 2. Hence, many techniques can be utilized for processing, such as equal gain combining (EGC), select the best (SB) or maximum ratio combining (MRC) [31]. In this work, each RYGB LDs-light unit is used to send a different data stream. Thus, SB scheme is used to find the pixel(s) that view each RYGB LDs-light unit. Subsequently, MRC is used to combine the signals from pixels that view each LDs-light unit as will be shown later. The effect of changing the location of the imaging receiver on the pixel's receiving area can be found in details in [11]. The optical receiver is located on the communication floor, which is 1 m above the floor (see Fig. 1 (a)).

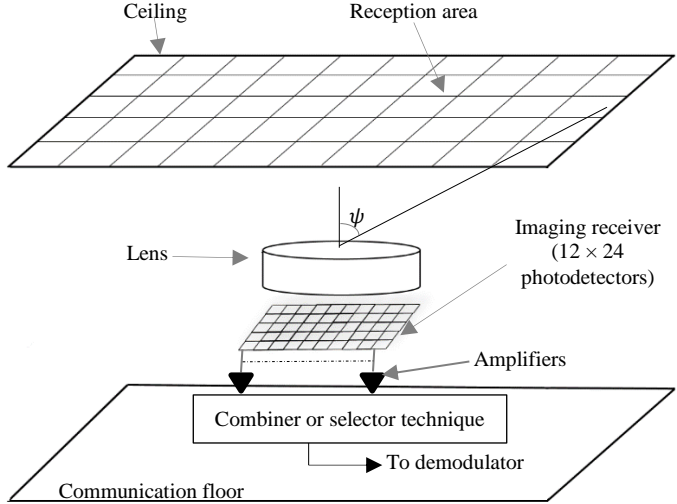

Fig. 2: Structure of the imaging receiver.

In this work, we consider a silicon photodetector. Thus, the pixel's bandwidth is given as [32]:

$$BW = \frac{1}{2 \pi R_l C_d} \quad (3)$$

where $R_l$ is the resistor of the load and $C_d$ is the pixel's capacitance. The relationship between the capacitance and the area of the pixel is [33]:

$$C_d = \frac{\varepsilon_0 \varepsilon_r A}{w} \quad (4)$$

here $A$ is the area of the pixel and $w$ is the thickness of the detector ($w = 100$ μm). To match the preamplifier and the subsequent circuits, the value of $R_l$ is selected to be 50 Ω [32]. Hence, each pixel has maximum bandwidth ~ 4.48 GHz and this can support a received data rate of up to 6.35 Gb/s as the bandwidth required is 0.7 times the bit rate when using OOK modulation [34]. Table I summarises the parameters utilized in the simulation.

TABLE I
SYSTEM PARAMETERS

| Parameters | Configurations | |
|---|---|---|
| **Rooms** | | |
| Room's length ($x$) | 8 m | |
| Room's width ($y$) | 4 m | |
| Room's height ($z$) | 3 m | |
| $\rho$ of xzWall | 0.8 | |
| $\rho$ of yzWall | 0.8 | |
| $\rho$ of xz op.Wall | 0.8 | |
| $\rho$ of yz op.Wall | 0.8 | |
| $\rho$ of Floor | 0.3 | |
| $\rho$ of Windows | 0 | |
| $\rho$ of Bookshelves | 0.4 | |
| Number of Bounces | 1 | 2 |
| Surface elements' number | 32000 | 2000 |
| $d_A$ | 5 cm × 5 cm | 20 cm × 20 cm |
| Lambertian emission order ($n$) | 1 | |
| Semi-angle | 60º | |
| **Light units** | | |
| Number of transmitters | 8 | |
| Places ($x$, $y$, $z$) m | (1, 1, 3), (1, 3, 3), (1, 5, 3), (1, 7, 3), (3, 1, 3), (3, 3, 3), (3, 5, 3), (3, 7, 3) | |
| RYGB LDs per unit | 9 (3 × 3) | |
| Interval between RYGB LDs | 0.03 m | |
| Optical power/RYGB LDs | 1.9 W | |
| Centre luminous intensity | 162 cd | |
| Lambertian emission order ($n$) | 0.65 | |
| Semi-angle (FOV) | 70º | |
| Azimuth of each light unit | 0º | |
| Elevation of the imaging receiver | 90º | |
| **Receiver** | | |
| Number of receivers | 1 | |
| Concentrator entrance area | $\frac{9\pi}{4}$ cm$^2$ | |
| Concentrator refractive index | 1.7 | |
| Concentrator acceptance angle | 65º | |
| FOV of each pixel | 4.7º | |
| Azimuth of the imaging receiver | 0º | |
| Elevation of the imaging receiver | 90º | |
| Receiver's area | 2 cm$^2$ | |
| Number of pixels | 288 | |
| Pixel's area | 0.694 mm$^2$ | |
| Responsivity | 0.4 A/W | |
| Receiver's bandwidth | 4 GHz | |

## IV. DESIGN OF PROPOSED VLC SYSTEM

In our proposed rooms, the RYGB LDs-light units (transmitters) were spatially separated to obtain a suitable illumination level. Moreover, each pixel of the imaging receiver was considered as a single photo-detector receiver with a small FOV. Thus, each transmitter was viewed by different pixel(s) of the imaging receiver and each transmitter was utilized to send a different data stream at a different data rate. We used SCM tones to (i) identify each light unit, (ii) determine the RYGB LDs-light units that can be used to transmit data (active light units), (iii) find the pixels that viewed these light units and (iv) calculate the level of CCI between the light units. The performance of each transmitter was evaluated based on the level of CCI and ISI due to the diffuse components (up to second order reflections).

### A. Distinguishing the RYGB LDs-light units

To distinguish the RYGB LDs-light units, each unit is given an ID. A preamble code or a packet header can be used to identify RYGB LDs-light units. The preamble code and the packet header are sequences of bits that can be added to the data frame of each RYGB LDs-light unit to distinguish them. For example, eight RYGB LDs-light units were used in our proposed rooms; hence, a sequence of three bits (000 - 111) may be assumed to identify each RYGB LDs-light unit. However, any error in decoding these sequence of bits used to identify each RYGB LDs-light unit would lead to a wrong decision at the receiver and loss of the data since the data is transmitted through the RYGB LDs-light units in parallel. Errors may occur in the header or preamble due to the RYGB LDs-light units emitting signals for broad coverage; hence, the ISI of the diffuse channel may cause an error when decoding the header and the preamble. In addition, intensity modulation with direct detection (IM/DD) are used in our VLC system, which means that each pixel of the imaging receiver responds in this case to all signals that are conveyed from all RYGB LDs-light units; thus, the CCI at the receiver can lead to an increase in the probability of error when decoding the header and the preamble.

Here, we propose SCM tones to identify each RYGB LDs-light unit, where these unmodulated tones are not affected by the channel dispersion [20]. Each RYGB LDs-light unit is multiplexed with a unique signal tone, which enables the receiver to easily distinguish the RYGB LDs-light units. A unique SCM tone ($f_1 - f_8$) was multiplexed into each RYGB LDs-light unit as can be seen in Fig. 3. It should be noted that the identification tones were used at the beginning of the communication to set up the connection between the light units and the imaging receiver (to find the RYGB LDs-light units that had good channel conditions with the optical receiver, to match these light units with the pixels of the imaging receiver and to find the level of the CCI) as will be explained further. The serial data were divided among only the RYGB LDs-light units that had a reliable connection with the imaging receiver (i.e., high channel dc gain, low level of CCI and low ISI); hence, the other RYGB LDs-light units were used for lighting only (no data transmitted through these units). In addition, each of the RYGB LDs-light units that were utilized to convey data, sent a different data stream at a different data rate while maintaining a target BER = $10^{-6}$.

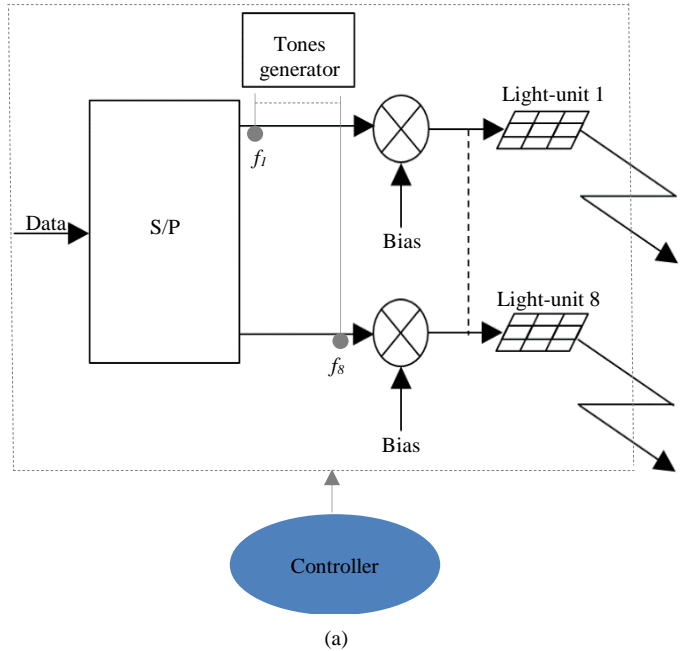

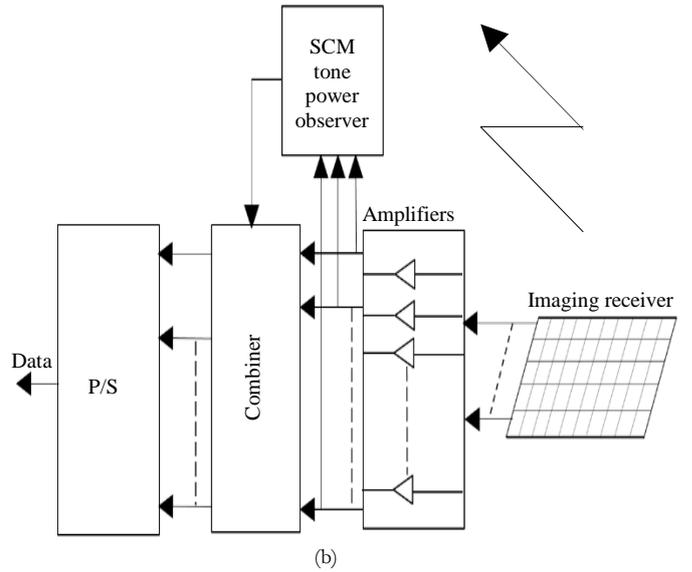

Fig. 3: Structure of the VLC system (a) transmitter and (b) receiver front-end.

### B. Connection setup between transmitters and imaging receiver

To find the RYGB LDs-light units that have a strong connection (units able to send a high data rate with a reliable connection; i.e. low BER) with the optical receiver, the controller arranges for the light units to send only the identification tones (no data transmission). At the receiver, the SCM tone identification system (see Fig. 3) was used to bandpass filter the received power due to each tone at each imaging receiver's pixel. The input current to RYGB LDs when including $m^{th}$ SCM tone is given as:

$$I_m(t) = I_{bais} + A_m sin(w_m t) \qquad (5)$$

where $I_{bais}$ is the bias current, $w_m = 2\pi f_m$, $f_m$ is the SCM tone frequency and $A_m$ is the amplitude of the SCM tone. The received optical power ($Pr_n$) of any SCM tone at the $n^{th}$ pixel of the imaging receiver due to $m^{th}$ light unit is given as:

$$Pr_n = \sum_{k=1}^{s} P_{LOS}^k + \sum_{i=1}^{l} P_{1st}^i + \sum_{d=1}^{y} P_{2nd}^d \qquad (6)$$

where $P_{LOS}$, $P_{1st}$ and $P_{2nd}$ are the received optical power from the $m^{th}$ RYGB light unit due to the LOS components, the first order reflection components and the second order reflection components, respectively, $s$ is the number of rays due to LOS components, $l$ is the number of rays due to the first order reflection components and $y$ is the number of rays due to the second order reflection components. Calculations of $P_{LOS}$, $P_{1st}$ and $P_{2nd}$ can be found in [23] and [35]. In this work, the $H$ matrix is formed by the received optical power ($Pr$) of each SCM tone at each pixel of the imaging receiver as:

$$H = \begin{bmatrix} Pr_{11} & \cdots & Pr_{1M} \\ \vdots & \ddots & \vdots \\ Pr_{N1} & \cdots & Pr_{NM} \end{bmatrix} \qquad (7)$$

where $N$ is the number of pixels and $M$ is the number of transmitters.

It should be noted that the channel frequency response of indoor VLC systems shows low pass characteristics in the electrical domain [36]. To ensure that the range of the frequencies that is given to the SCM tones is located where the attenuation of the channel is low, we calculated the 3-dB bandwidth of the indoor VLC channel. In our system, however, each RYGB light unit has a different channel frequency response as each RYGB light unit sends a different data stream. Therefore, we calculated the 3-dB channel bandwidth of each light unit when the optical receiver was placed at the centre of room A (2 m, 4 m, 1 m) as the imaging receiver sees all light units and the distances between the light units and the optical receiver become maximum at this location. The channel frequency response of each RYGB light unit is given as:

$$H(f_m) = \int_0^\infty h_m(t)e^{-j2\pi f_m t}\, dt \qquad (8)$$

here $h(t)$ is the indoor channel's impulse response (we obtained $h(t)$ in a fashion similar to that in [35] for each RYGB light unit).

Table II shows the 3-dB channel bandwidth due to each RYGB light unit when the optical receiver was placed at the centre of room A. It can be seen that RYGB light units 1, 2, 7 and 8 have the same 3-dB channel bandwidth and RYGB light units 3, 4, 5 and 6 have similar 3-dB channel bandwidth. This is attributed to the symmetry of room A. As seen in Table II, the lowest 3-dB channel bandwidth is 1.93 GHz. Therefore, the range of the frequencies chosen for the SCM tones was 500 MHz to 920 MHz with a guard interval of 60 MHz, which also enables the use of low cost electronic components.

TABLE II
3-dB CHANNEL BANDWIDTH OF RYGB LIGHT UNITS WHEN THE OPTICAL RECEIVER WAS LOCATED AT THE CENTER OF ROOM A.

| RYGB light units | Unit 1 | Unit 2 | Unit 3 | Unit 4 | Unit 5 | Unit 6 | Unit 7 | Unit 8 |
|---|---|---|---|---|---|---|---|---|
| 3-dB channel bandwidth (GHz) | 1.93 | 1.93 | 5.37 | 5.37 | 5.37 | 5.37 | 1.93 | 1.93 |

Each pixel of the imaging receiver receives optical power from all light units (either due to LOS components or due to reflection components). Accordingly, eight bandpass filters (BPFs), which have centre frequencies equal to the frequencies of the SCM tones, are used in the SCM tone identification system to separate the SCM tones at each pixel of the imaging receiver as shown in Fig 4. It should be noted that the SCM tone identification system is used at the beginning of the communication to set up the connection between the light units and the optical receiver. Therefore, during just this monitoring time the received signals from light units enter the BPFs. The output of each BPF is either a desired SCM tone (the SCM tone that has LOS component with the imaging receiver) plus noise or an undesired SCM tone (the SCM tone that has no LOS component with the imaging receiver) plus noise. Therefore, to decide the output of the BPF, an optimum threshold should be determined.

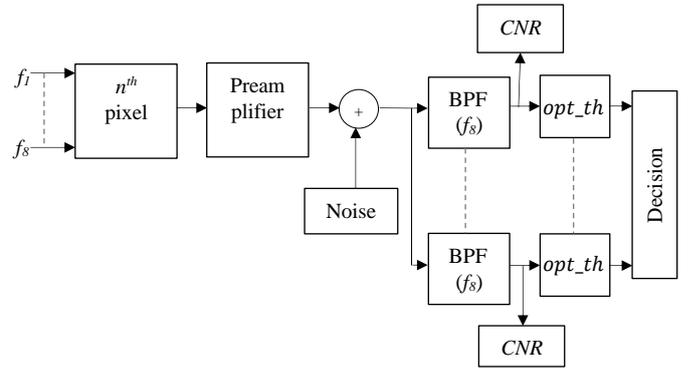

Fig. 4: Architecture of the SCM tone identification system.

To identify the output electrical current ($z$) of each BPF, we hypothesize two situations:
Hypothesis 1 ($H1$): $z =$ undesired SCM tone plus noise ($b + n$).
Hypothesis 2 ($H2$): $z =$ desired SCM tone plus noise ($a + n$).
The total noise ($n$) that is seen by each SCM tone is white Gaussian zero mean, with total standard deviation $\sigma_t$. It is generated by shot noise due to the background light, signal shot niose and preamplifier thermal niose. The total standard deviation of the noise ($\sigma_t$) can thus be written as [37]:

$$\sigma_t = \sqrt{\sigma_{bn}^2 + \sigma_s^2 + \sigma_{pr}^2} \qquad (9)$$

where $\sigma_{bn}$ is the ambient shot noise, $\sigma_s$ is the shot noise related with the SCM tone and $\sigma_{pr}$ is the pre-amplifier thermal noise. The background light shot niose ($\sigma_{bn}$) is given as:

$$\sigma_{bn} = \sqrt{2qAI_{bn}BW_{BPF}} \qquad (10)$$

here $q$ is the electron charge, $I_{bn}$ is the background photocurrent, which is induced due to the light from the sky and background light sources ($I_{bn}= 10^{-3}$ A/cm$^2$) [22], and $BW_{BPF}$ is the bandwidth of the BPF, which was selected to be equal to 4 MHz (following BER optimisation using the results in this section), to reduce the noise seen by each SCM tone. The shot noise induced by the SCM tones is expressed as [38]:

$$\sigma_s = \sqrt{2qRPr_n BW_{BPF}} \qquad (11)$$

In this paper, the p-i-n FET receiver designed in [39] was used. The input noise current for this receiver is equal to 4.5 pA/$\sqrt{Hz}$.

The values of $a$ and $b$ (the electrical currents of the desired SCM tone and the undesired SCM tone) depend on the distance between the optical receiver and the light units. Thus, $a$ and $b$ are random variables whose distribution we determined using our ray tracing modelling tool by considering 1000 random locations of the optical receiver on the communication floor of room A. Due to the symmetry of room A, we considered light unit 1 as the desired transmitter and light unit 2 as the interfering source. At each location of the receiver, we obtained the electrical current of $a$ and $b$ generated by the desired SCM tone and the undesired SCM tone. Fig. 5 and Fig. 6 show the histogram and the curve fitting of the desired SCM tone and the undesired SCM tone, respectively.

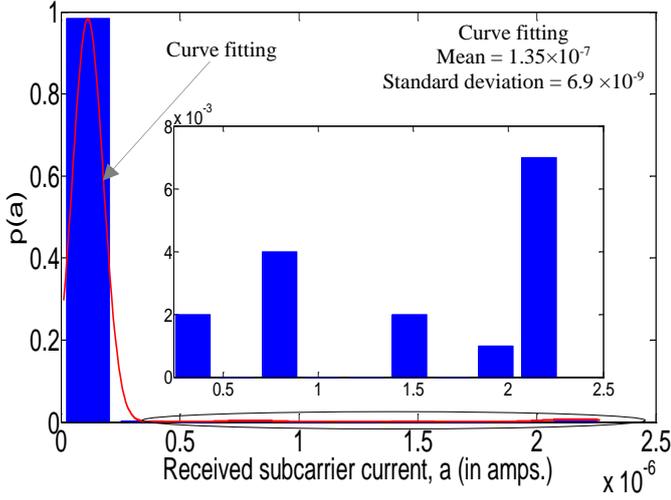

Fig. 5: Histogram and the curve fitting of the desired SCM tone.

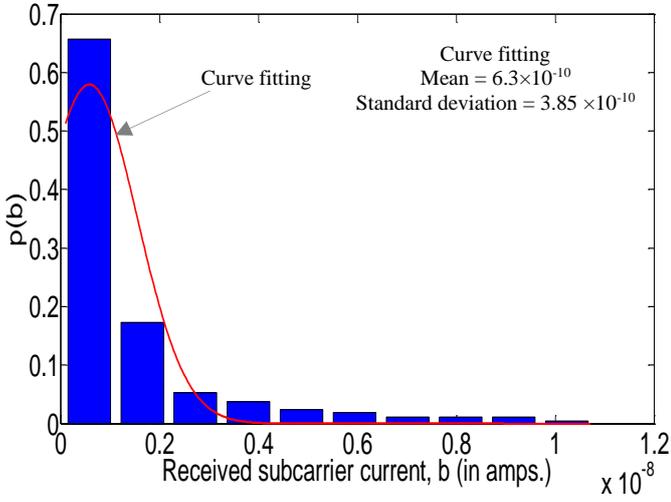

Fig. 6: Histogram and the curve fitting of the undesired SCM tone.

From the curve fitting, the normalized probability density functions (pdfs) of $a$ ($p(a)$) and $b$ ($p(b)$) are given as:

$$p(a) = \frac{1}{\sqrt{2\pi}\,\sigma_{ds}} e^{-\left(\frac{a-m_{ds}}{\sqrt{2}\sigma_{ds}}\right)^2} \qquad (12)$$

and

$$p(b) = \frac{1}{\sqrt{2\pi}\,\sigma_{us}} e^{-\left(\frac{b-m_{us}}{\sqrt{2}\sigma_{us}}\right)^2} \qquad (13)$$

where we assumed a Gaussian distribution for the received power which is reasonable given the multiple reflecting surfaces and the observed results. Here, $m_{ds}$ and $\sigma_{ds}$ are the mean value and the standard deviation of the desired SCM tone, respectively and $m_{us}$ and $\sigma_{us}$ are the mean value and the standard deviation of the undesired SCM tone, respectively.

We can write the pdf's of $z$ given hypotheses $H1$ and $H2$, respectively. Under hypothesis $H1$, $z$ is the convolution of the noise pdf and the undesired SCM tone pdf and is given as:

$$fz(z|H1) = p(b) \otimes p(n) \qquad (14)$$

Solving (14), $fz(z|H1)$ can be written as:

$$fz(z|H1) = \frac{1}{\sqrt{2\pi(\sigma_{us}^2 + \sigma_t^2)}} e^{-\left(\frac{z-m_{us}}{\sqrt{2(\sigma_{us}^2+\sigma_t^2)}}\right)^2} \qquad (15)$$

Under hypothesis $H2$, $z$ is the convolution of the noise pdf and the desired SCM tone pdf and is given as:

$$fz(z|H2) = \frac{1}{\sqrt{2\pi(\sigma_{ds}^2 + \sigma_t^2)}} e^{-\left(\frac{z-m_{ds}}{\sqrt{2(\sigma_{ds}^2+\sigma_t^2)}}\right)^2} \qquad (16)$$

Applying likelihood ratio to equations (15) and (16), we get:

$$\frac{fz(z|H2)}{fz(z|H1)} \underset{H1}{\overset{H2}{\gtrless}} 1$$

$$= \frac{\frac{1}{\sqrt{2\pi(\sigma_{ds}^2 + \sigma_t^2)}} e^{-\left(\frac{z-m_{ds}}{\sqrt{2(\sigma_{ds}^2+\sigma_t^2)}}\right)^2}}{\frac{1}{\sqrt{2\pi(\sigma_{us}^2 + \sigma_t^2)}} e^{-\left(\frac{z-m_{us}}{\sqrt{2(\sigma_{us}^2+\sigma_t^2)}}\right)^2}} \underset{H1}{\overset{H2}{\gtrless}} 1 \qquad (17)$$

Solving equation (17) we get:

$$z \underset{H1}{\overset{H2}{\gtrless}} \frac{1}{\sigma_{ds}^2 - \sigma_{us}^2}\left( m_{us}(\sigma_{ds}^2 + \sigma_t^2) - m_{ds}(\sigma_{us}^2 + \sigma_t^2) \right.$$
$$+ \sqrt{\begin{pmatrix} \sigma_{ds}^2\sigma_{us}^4 + \sigma_t^4\sigma_{us}^2 + \sigma_t^2\sigma_{us}^4 \\ -\sigma_{ds}^4\sigma_t^2 - \sigma_{ds}^4\sigma_{us}^2 - \sigma_{ds}^2\sigma_t^4 \end{pmatrix}\ln\left(\frac{\sigma_{us}^2 + \sigma_t^2}{\sigma_{ds}^2 + \sigma_t^2}\right) + \\ (m_{ds} - m_{us})^2(\sigma_t^2\sigma_{ds}^2 + \sigma_{ds}^2\sigma_{us}^2 + \sigma_t^4 + \sigma_t^2\sigma_{us}^2)}\left.\right)$$
$$= z \underset{H1}{\overset{H2}{\gtrless}} opt\_th \qquad (18)$$

It should be noted that when the $\sigma_{ds}$, $\sigma_{us}$ and $m_{us}$ are very small compared with $m_{ds}$, the optimum threshold ($opt\_th$) $\sim \frac{m_{ds}}{2}$. Fig. 7 shows the relationship between $\sigma_{ds}$ and $z$ using (18). As can be seen in Fig. 7, the optimum threshold $\sim \frac{m_{ds}}{2}$ when $\sigma_{ds} < 10^{-9}$.

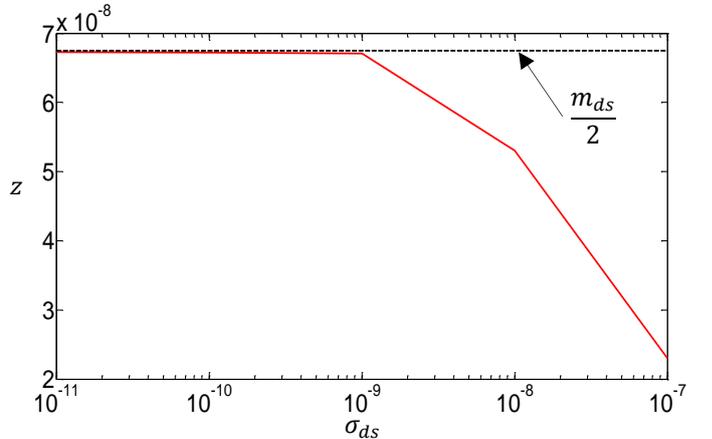

Fig. 7: Relationship between the standard deviation of the desired SCM tone ($\sigma_{ds}$) and the optimum threshold ($opt\_th$).

It should be noted that the standard deviation of the desired SCM tone ($\sigma_{ds}$) depends on the size of the proposed rooms, light units' distribution, distance between transmitters and the receiver and the FOV of the receiver. Thus, changing these parameters changes the optimum threshold value.

The probability of detection of the desired SCM tone, which is the probability of correct decision of the desired SCM tone ($Pcds$) is given as:

$$Pcds = \int_{opt\_th}^{\infty} fz(z|H2)dz =$$
$$\int_{opt\_th}^{\infty} \frac{1}{\sqrt{2\pi(\sigma_{ds}^2+\sigma_t^2)}} e^{-\left(\frac{z-m_{ds}}{\sqrt{2(\sigma_{ds}^2+\sigma_t^2)}}\right)^2} dz \qquad (19)$$

Consequently, the probability of detection of the undesired SCM tone, which is the probability of false alarm of the undesired SCM ($Pfus$) tone (i.e. the undesired SCM tone considered as the desired SCM tone) is given as:

$$Pfus = \int_{opt_{th}}^{\infty} fz(z|H1)dz =$$
$$\frac{1}{\sqrt{2\pi(\sigma_{us}^2+\sigma_t^2)}} e^{-\left(\frac{z-m_{us}}{\sqrt{2(\sigma_{us}^2+\sigma_t^2)}}\right)^2} dz \qquad (20)$$

Hence, the probability of not assigning the undesired SCM tone to a user, which is the probability of the correct decision in the undesired SCM tone ($Pcus$) is:

$$Pcus = 1 - Pfus \qquad (21)$$

The overall probability of the SCM identification system to make a correct decision ($Pcd$) is:

$$Pcd = Pcds \, (Pcus)^{M-1} \qquad (22)$$

Consequently, the probability of being wrong in the decision ($Pwd$) is $1 - Pcd$. In our system, and for the given set of parameters in this paper, $Pwd$ is $1.45 \times 10^{-10}$, which can be ignored as each RYGB light unit sends data with a target BER = $10^{-6}$. Therefore, our SCM identification system is able to associate each pixel(s) of the imaging receiver with a light unit at a low "association probability of error, $Pwd$".

We next use the SCM tones as a simple mechanism to estimate the CCI in our system. The CCI level is defined as the total received power at the $n^{th}$ pixel, except for the received optical power of the desired signal (SCM tone here). For example, if the desired tone is $f_m$ (identification tone of $m^{th}$ light unit), the electrical power of this tone ($S_{n,m}$) and the level of CCI due to the other tones at $n^{th}$ pixel of the imaging receiver can be given as:

$$S_{n,m} = \frac{(R\, Pr_{n,m})^2}{2} \text{ and } I_{n,k} = \sum_{\substack{k=1 \\ k \neq m}}^{M} \frac{(R\, Pr_{n,k})^2}{2},$$
$$n \in [1, 2 \ldots N], \ m \in [1, 2 \ldots M] \qquad (23)$$

The carrier to noise ($CNR$) ratio of the $m^{th}$ tone at the $n^{th}$ pixel of the imaging receiver is given as:

$$CNR_{n,m} = \frac{(R\, Pr_{nm})^2}{2\, \sigma_t^2} \qquad (24)$$

A few pixels of the imaging receiver are used to receive the parallel data. In addition, based on the channel condition of each RYGB light unit with the optical receiver, not all RYGB light units were used to transmit the data. Therefore, to exclude the unwanted pixels (pixels that collected very low optical power from the RYGB light units) and to find the RYGB light unit that has the ability to send data, we set up a threshold level for each pixel. We defined the factor $CNR/I$ as the threshold level:

$$\left(\frac{CNR}{I}\right)_{n,m} = \frac{(R\, Pr_{nm})^2}{2(\sigma_t^2 + I)} \qquad (25)$$

Hence, if any RYGB LDs-light unit has $CNR/I > 13.6$ dB (BER = $10^{-6}$) at any pixel of the imaging receiver, the optical receiver informs the controller to consider this light unit to transmit data, else switch OFF this RYGB LDs-light unit illumination only. The feedback signal can be a low data rate infrared channel similar to the one; we considered [40]. In addition, if this RYGB light unit is used for data communication, the SCM tone identification system informs the combiner (see Fig. 3) of the pixel(s) that viewed this RYGB LDs-light unit. Accordingly, the combiner selects the pixel(s) (by using SB scheme) of the imaging receiver that have a good connection with the light units and excludes the other pixels. It should be noted in our system, we used MRC to combine the signals from the pixels that see a given LDs-light unit and therefore determined the aggregate data. The signal to interference and noise ratio (SINR) was used to evaluate the performance of the data channels. Thus, 13.6 dB was chosen, which gives BER = $10^{-6}$ for OOK modulation. The signal to interference and noise ratio (SINR) was used to evaluate the performance of the data channels. Thus, 13.6 dB was chosen, which gives BER = $10^{-6}$ for OOK modulation.

*C. Performance analysis of the proposed system*

To calculate the maximum data rate that can be sent through each light unit of the active RYGB LDs-light units, the ISI due to multipath dispersion of the data signal should be determined. The delay spread ($D$) associated with a given impulse response can be given as [41]:

$$D = \sqrt{\frac{\sum(t_i - \mu)^2 P_{ri}^2}{\sum P_{ri}^2}} \qquad (26)$$

here $t_i$, $P_{ri}$ and $\mu$ are the delay time, the received optical power and the mean delay, respectively. The mean delay is given as:

$$\mu = \frac{\sum t_i P_{ri}^2}{\sum P_{ri}^2} \qquad (27)$$

When using OOK, the relationship between the $BER$ and the $SINR$ is given as [42]:

$$BER = Q(\sqrt{SINR}) \quad (28)$$

here $Q(x) = \frac{\int_x^\infty e^{-z^2/2} dz}{\sqrt{2\pi}}$. The SINR can be expressed as [42], [43]:

$$SINR = \frac{R^2(P_{s1} - P_{s0})^2}{\sigma_{dt}^2 + I} \quad (29)$$

where $P_{s1}$ and $P_{s0}$ are the received powers of logic 1, and logic 0, respectively and $\sigma_{dt}$ is the total noise associated with the received data. It should be noted that to calculate $\sigma_{dt}$, we considered the receiver bandwidth, which is 4 GHz. The received power due to logic 1, ($P_{s1}$) and logic 0 ($P_{s0}$) can be determined from the transmitted signal and the channel impulse response.

In some locations of the optical receiver, more than one pixel views the same light unit. In this case, an MRC scheme was used to combine the signals from these pixels, and calculate the SINR as:

$$SINR_{MRC} = \sum_{i=1}^{J} \left( \frac{R^2(P_{s1i} - P_{s0i})^2}{\sigma_{dti}^2 + I_i} \right) \quad (30)$$

here $J$ is the number of pixels that see the same light unit.

## V. RESULTS OF THE SIMULATION

The performance of our VLC system in room A (empty room) and the room B (realistic room) was assessed. A simulation tool like the one we reported in [44]-[48] was used. A MATLAB program is utilized for the analyses in this paper. The performance of our VLC system was examined at many coordinates on rooms' communication floor. For room A, we obtained the results when the mobile user moved at $x = 1$ m and $x = 2$ m and along the $y$-axis. This is attributed to the symmetry of room A. For room B, however, the results were evaluated at $x = 1$ m, $x = 2$ m and $x = 3$ m and along the $y$-axis.

The RYGB LDs-light units delay spread when the optical receiver was located along the $y$-axis and at $x = 1$ m and $x = 2$ m in room A is shown in Table III. Observe that the delay spread is affected by the delay time and the optical received power of the rays (see equation 26). Thus, at each location of the imaging receiver, the RYGB LDs-light units near the receiver have delay spread lower than the delay spread of further LDs-light units as shown in Table III.

Table IV shows the active RYGB LDs-light units that were utilized to convey the data and the maximum data rate of each active unit (data rate that can be transmitted with SINR = 13.6 dB, BER = $10^{-6}$) when the optical receiver was placed at the corner (1 m, 1 m, 1 m) and the centre (2 m, 4 m 1 m) of room A. The performance of each active light unit was determined based on the channel dc gain, level of CCI due to other light units, and ISI due to diffusing reflections. As can be seen, three RYGB LDs-light units (unit 1, unit 2 and unit 3) were used to carry the data when the optical receiver was placed at the corner of room A, whereas four units (unit 3, unit 4, unit 5 and unit 6) were utilized to send the data when the receiver was located at the centre of room A. This is related to the poor performance of other light units (light units that were used for lighting only). In other words, the levels of the CCI and the ISI were high for these light units. In addition, it can be seen that by placing the optical receiver at the centre of the room, all active units sent a different data stream at the same rate, and this is attributed to the symmetry of room A (see Fig. 1 (a)).

TABLE IV
ACTIVE RYGB LDS-LIGHT UNITS CONVEYING DATA WHEN THE RECEIVER WAS LOCATED AT (1 m, 1 m, 1 m) AND AT (2 m, 4 m 1 m) IN ROOM A.

| Receiver is located at room corner (1 m, 1 m, 1 m) | | | Receiver is located at room center (2 m, 4 m, 1 m) | | |
|---|---|---|---|---|---|
| RYGB LDs-light units | Status | Data rate (Gb/s) | RYGB LDs-light units | Status | Data rate (Gb/s) |
| Transmitter1 | ON | 4 | Transmitter1 | OFF | - |
| Transmitter2 | ON | 2.4 | Transmitter2 | OFF | - |
| Transmitter3 | ON | 1.75 | Transmitter3 | ON | 2.42 |
| Transmitter4 | OFF | - | Transmitter4 | ON | 2.42 |
| Transmitter5 | OFF | - | Transmitter5 | ON | 2.42 |
| Transmitter6 | OFF | - | Transmitter6 | ON | 2.42 |
| Transmitter7 | OFF | - | Transmitter7 | OFF | - |
| Transmitter8 | OFF | - | Transmitter8 | OFF | - |

Fig. 8 illustrates the aggregate data rate of our VLC system when the receiver is placed at many positions on the communication floor of room A at $x = 1$ m and $x = 2$ m and along the $y$-axis. The aggregate data rate was obtained at BER = $10^{-6}$ (SINR = 13.6 dB) for the received data for each active RYGB LDs-light unit. As shown, the lowest aggregate data rate (8.05 Gb/s) was achieved when the optical receiver is placed at the corner of room A (1 m, 1 m, 1 m). This is due to some of the light units being out of the FOV of the imaging receiver at this position. The maximum aggregate data rate (8.95 Gb/s) was obtained when the optical receiver was positioned at the room centre (2 m, 4 m, 1 m), and this is attributed to the imaging receiver being able to view more transmitters with good performance (low CCI level and low ISI) at this location. The maximum data rate transmitted through each active RYGB LDs-light unit does not exceed the data rate supported by each the pixel of the imaging receiver.

TABLE III
DELAY SPREAD OF EACH RYGB LDs-LIGHT UNIT WHEN THE OPTICAL RECEIVER WAS LOCATED AT MANY LOCATIONS IN ROOM A.

| RYGB LDs-Light units | Delay spread (ns) | | | | | | | | | | | | | |
|---|---|---|---|---|---|---|---|---|---|---|---|---|---|---|
| | Receiver locations (m) | | | | | | | Receiver locations (m) | | | | | | |
| | 1,1,1 | 1,2,1 | 1,3,1 | 1,4,1 | 1,5,1 | 1,6,1 | 1,7,1 | 2,1,1 | 2,2,1 | 2,3,1 | 2,4,1 | 2,5,1 | 2,6,1 | 2,7,1 |
| Transmitter1 | 0.021 | 0.027 | 0.054 | 0.4 | 0.7 | 4.2 | 4.4 | 0.05 | 0.053 | 0.41 | 0.52 | 3.1 | 3.1 | 3.2 |
| Transmitter2 | 0.063 | 1.65 | 1.9 | 2.8 | 4.4 | 4.8 | 5.1 | 0.051 | 0.052 | 0.4 | 0.52 | 1.03 | 2.9 | 3.3 |
| Transmitter3 | 0.05 | 0.028 | 0.022 | 0.035 | 0.05 | 0.08 | 0.4 | 0.41 | 0.05 | 0.03 | 0.031 | 0.49 | 0.08 | 1.2 |
| Transmitter4 | 1.57 | 2.11 | 0.04 | 2.1 | 1.7 | 1.8 | 2.1 | 0.41 | 0.05 | 0.03 | 0.031 | 0.47 | 0.081 | 1.3 |
| Transmitter5 | 0.3 | 0.08 | 0.051 | 0.033 | 0.021 | 0.028 | 0.05 | 1.2 | 0.08 | 0.5 | 0.031 | 0.029 | 0.05 | 0.41 |
| Transmitter6 | 2.1 | 1.8 | 1.6 | 2.2 | 0.04 | 2 | 1.6 | 1.3 | 0.081 | 0.48 | 0.031 | 0.03 | 0.05 | 0.4 |
| Transmitter7 | 4.5 | 4.1 | 0.6 | 2.8 | 0.053 | 0.027 | 0.021 | 3.1 | 2.9 | 1.06 | 0.51 | 0.3 | 0.05 | 0.048 |
| Transmitter8 | 5.2 | 4.8 | 4.5 | 0.39 | 2.1 | 1.7 | 0.06 | 3.2 | 2.9 | 1.06 | 0.52 | 0.41 | 0.049 | 0.045 |

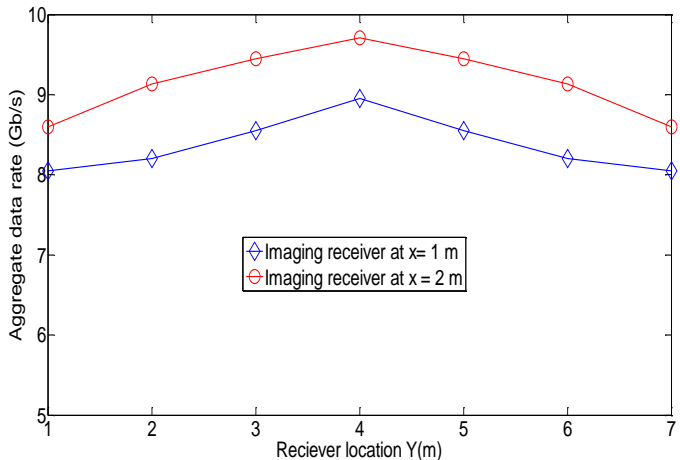

Fig 8: An aggregate data rate of the proposed indoor system at different locations of the optical receiver in room A with BER = $10^{-6}$ (SINR = 13.6 dB).

In order to consider the effect of the obstacles on our proposed VLC system, we extended the analysis to room B (the realistic room). In room B, we account for the effects of signal blockage, due to small cubicles, a door, windows, and furniture and multipath propagation (see Fig. 1 (b)). Due to the asymmetry of room B, the mobile receiver was considered at different positions on the lines $x = 1$ m, $x = 2$ m and $x = 3$ m along the $y$-axis.

The aggregate data rates of the system when the mobile receiver is placed at many locations along the $y$-axis at $x = 1$ m, $x = 2$ m and $x = 3$ m in room B are shown in Fig. 9. In general, the performance of the system was better when the mobile receiver moved along $x = 2$ m. This is because of the optical receiver viewing more light units along the $x = 2$ m when compared with $x = 1$ m and $x = 3$ m. As can be seen, the aggregate data rate along $x = 1$ m was better than the aggregate data rate along $x = 3$ m. As along $x = 1$ m, the nearest walls ($x = 0$ and $y = 0$) to the optical receiver are windows and these windows reflection coefficients are zero; whereas, along $x = 3$ m, the nearest walls (wall $x = 4$ m and wall $y = 8$ m) to the optical receiver are covered by bookshelves, which have reflection coefficients of 0.4. Therefore, the CCI level and ISI due to reflection components were higher along $x = 3$ m compared with $x = 1$ m, which led to a decrease in the aggregate data rate along $x = 3$ m.

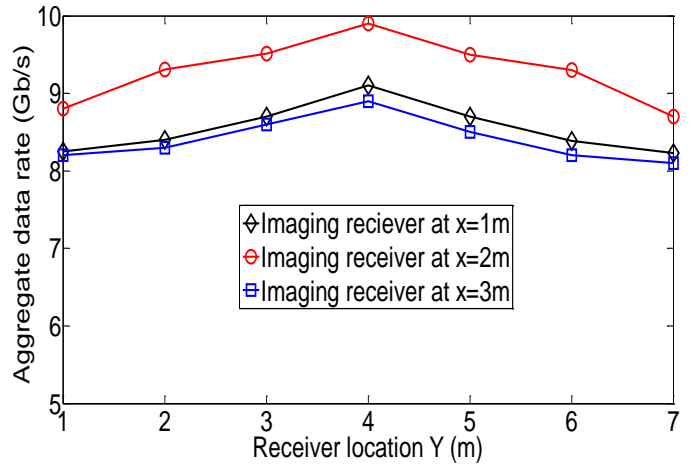

Fig. 9: Aggregate data rate of the proposed indoor VLC system at different locations of the optical receiver in room B with BER = $10^{-6}$ (SINR = 13.6 dB).

## VI. CONCLUSIONS

In this paper, we introduced an indoor VLC system that can offer a data rate of 8 Gb/s with BER = $10^{-6}$ (SINR = 13.6 dB) using simple OOK modulation. The system used parallel data transmission sent from multiple RYGB LDs-light units located on the room's ceiling to provide suitable level of lighting. An imaging receiver was used in this work to benefit from spatial multiplexing and to mitigate the effects of ISI. SCM tones were proposed in this paper to identify each RYGB LDs-light unit and to find the pixel(s) that received the signals from these light units. Additionally, these tones were used to calculate the CCI level between the transmitters. Based on the performance (CCI level and ISI) of each RYGB LDs-light unit, the serial data were divided (only) between the light units that were able to send a high data rate with a strong connection. Our proposed system was investigated in two different rooms (empty and realistic rooms) while considering the effect of CCI and diffuse reflections (up to second order).